\documentclass[reprint,twocolumn,superscriptaddress,showpacs,nofootinbib,notitlepage,preprintnumbers]{revtex4-2}

\usepackage{graphicx}
\usepackage{hyperref}
\usepackage[dvipsnames]{xcolor}
\usepackage{amsmath,amsthm,amssymb}
\usepackage{comment}
\usepackage{algorithm,algpseudocode}
\usepackage{enumerate}

\newcommand{\todo}{\colorbox{pink}{\textsc{Todo}}}

\renewcommand{\Re}{\mathrm{Re}\,}
\renewcommand{\Im}{\mathrm{Im}\,}

\DeclareMathOperator{\Tr}{Tr}

\newtheorem{definition}{Definition}
\newtheorem*{conjecture*}{Conjecture}

\begin{document}
\preprint{LA-UR-24-28907}

\title{Model-free spectral reconstruction via Lagrange duality}

\author{Scott Lawrence}
\email{srlawrence@lanl.gov}
\affiliation{Los Alamos National Laboratory Theoretical Division T-2, Los Alamos, NM 87545, USA}
\date{\today}

\begin{abstract}
	Various physical quantities---including real-time response, inclusive cross-sections, and decay rates---may not be directly determined from Euclidean correlators. They are, however, easily determined from the spectral density, motivating the task of estimating a spectral density from a Euclidean correlator.
	This spectral reconstruction problem can be written as an ill-posed inverse Laplace transform; incorporating positivity constraints allows one to obtain finite-sized bounds on the region of spectral density functions consistent with the Euclidean data. Expressing the reconstruction problem as a convex optimization problem and exploiting Lagrange duality, bounds on arbitrary integrals of the spectral density can be efficiently obtained from Euclidean data. This paper applies this approach to reconstructing a smeared spectral density and determining smeared real-time evolution. Bounds of this form are information-theoretically complete, in the sense that for any point within the bounds one may find an associated spectral density consistent with both the available Euclidean data and positivity.
\end{abstract}

\maketitle

\section{Introduction}\label{sec:introduction}
Lattice Monte Carlo methods provide practical, nonperturbative calculations of a restricted set of observables in a restricted set of quantum systems. Excluded from the set of systems are, most famously, relativistic systems at finite fermion density and condensed-matter systems away from half-filling, due to the fermion sign problem~\cite{PhysRevB.41.9301}. The set of efficiently computable observables is typically considered to exclude all those that cannot be expressed in terms of the Euclidean-time correlator, including (dynamical) transport coefficients and hadronic decay rates. This paper is chiefly concerned with the task of determining what information regarding such observables is available in Euclidean correlators, and how it may be extracted.

An obvious alternative to extracting real-time information from the Euclidean path integral is to work instead with a path integral in real (Minkowski) time. Working directly with a lattice path integral in real time one encounters a sign problem, similar in some ways to the fermion sign problem, which may be mitigated via contour deformations~\cite{Alexandru:2016gsd,Alexandru:2017lqr,Lawrence:2021izu,Lawrence:2024pjg} or entirely avoided via complex Langevin~\cite{Lampl:2023xpb,Boguslavski:2022dee,Alvestad:2021hsi}. The former approach has been applied successfully only to quantum mechanics and one-dimensional lattice field theories, and is now known to have an exponential residual sign problem in the general case~\cite{Lawrence:2023sfc}. The latter approach has no residual sign problem, but it is difficult in practice to guarantee convergence to the physical answer~\cite{Nagata:2016vkn,Nagata:2018net,Seiler:2023kes}. The task of determining real-time behavior of quantum systems has also motivated the development of quantum algorithms for a variety of systems~\cite{feynman2018simulating,Cohen:2021imf}.

Near equilibrium, the real-time response of a thermalized system to a perturbation is characterized by real-time correlation functions in the thermal system. In principle these correlators may be determined by analytic continuation from the Euclidean correlators obtained on the lattice---or equivalently, by solving the \emph{inverse problem} to determine the spectral density function from the Euclidean data. Such inverse problems have a central role in the determination of transport coefficients from lattice data~\cite{Meyer:2007ic,Meyer:2007dy}. In addition, spectral reconstruction from lattice QCD is central to determining the $R$-ratio~\cite{ExtendedTwistedMassCollaborationETMC:2022sta}, and decay and transition rates~\cite{Hansen:2017mnd}. Inverse problems further appear in the determination of parton distributions~\cite{Ji:2013dva} and sphaleron rates~\cite{Bonanno:2023gwz}

Applications of the same methods extend well outside lattice field theory. An early approach to the inverse problem, today known as the Backus-Gilbert method, was first outlined in geophysics~\cite{backus1968resolving,backus1970uniqueness}. More recent developments provide bounds on arbitrary integrals of the spectral density, given Euclidean data free of statistical errors~\cite{Hansen:2019idp}. Other approaches make use of analytic structure~\cite{Bergamaschi:2023xzx,fei2021nevanlinna}, Pad\'e fits~\cite{Tripolt:2018xeo}  or machine learning methods to regularize the otherwise ill-posed inverse problem~\cite{Chen:2021giw,Wang:2021jou,fournier2020artificial,Kades:2019wtd,Horak:2021syv}.

The method described in this work is centered around a single observation: the space of spectral density functions, constrained both by statistical consistency with the observed Euclidean correlator and by the requirement that the spectral density be non-negative, is convex.  As a result, standard techniques from convex optimization allow us to efficiently explore this space, in particular providing rigorous upper and lower bounds for any linear functional of the spectral density (e.g.~smeared real-time correlators). These bounds are tight, in the sense that for any value inside these bounds, a non-negative spectral density may be found, consistent with the provided Euclidean data, on which the linear functional attains that value.

Two main technical tools from convex analysis are exploited in this work: \emph{Lagrange duality} and \emph{interior-point methods}. In the context of this work, Lagrange duality allows us to replace an optimization over the infinite-dimensional space of spectral density functions with an optimization over a finite-dimensional space of Lagrange multipliers. Interior-point methods are a class of algorithms for solving convex optimization problems; that is, for finding the minimum (or maximum) of a convex function on a convex space. Related approaches have proven effective in constraining scaling dimensions in CFTs~\cite{Kos:2016ysd,Poland:2018epd}, eigenstates of various quantum mechanical systems~\cite{Han:2020bkb,Berenstein:2021dyf,Berenstein:2021loy,Berenstein:2022ygg,Berenstein:2022unr,Berenstein:2024ebf}, and zero-temperature physics of lattice quantum field theories~\cite{barthel2012solving,Lawrence:2021msm,Lawrence:2022vsb}.

The remainder of this paper is organized as follows. Section~\ref{sec:spectral} defines the spectral density function and relates it to the Euclidean and real-time correlators, and briefly discusses the nature of the ill-posed inversion (or equivalently, analytic continuation) problem. The construction of convex programs for performing the inversion---that is, for bounding certain integrals of the spectral density---is detailed in Section~\ref{sec:convex}. With little modification the resulting algorithm is applied to compute smeared spectral densities in Section~\ref{sec:density} and smeared real-time correlators in Section~\ref{sec:realtime}. Higher precision can be obtained in the real-time response at lower temperatures, and Section~\ref{sec:cold} exploits this fact to obtain nontrivial constraints on the real-time behavior of scalar field theory in $2+1$ dimensions.
Finally, possible further applications, and limitations, of this method are discussed in Section~\ref{sec:discussion}. (Further technical and implementation details are given in appendices.)

The code used to perform positivity-constrained inversion has been made available online~\cite{scamp}.

\section{Spectral density functions}\label{sec:spectral}
Spectral density functions appear in many contexts. In this paper we will focus on out-of-equilibrium dynamics as our primary motivation.

The minimal departure of a quantum system from thermal equilibrium is described by \emph{linear response}. Concretely, we will consider the time-dependence of the expectation of a Hermitian operator $\mathcal O$ in the presence of a Hamiltonian of the form
\begin{equation}
	H(t) = H_0 + \epsilon \delta(t) \mathcal O
	\text.
\end{equation}
The state at times $t < 0$ is taken to be a thermal state $\rho = e^{-\beta H_0}$ of the unperturbed Hamiltonian. Expanding to linear order in $\epsilon$ we find
\begin{equation}
	\langle \mathcal O(t) \rangle = \langle \mathcal O(0)\rangle_0 + 2 \epsilon \Im \langle \mathcal O(t) \mathcal O(0) \rangle_0
	\text,
\end{equation}
where the expectations on the right-hand side are taken with respect to the unperturbed thermal state. We shall henceforth define the real-time (retarded) correlator $G(t) = \Im \langle \mathcal O(t) \mathcal O(0)\rangle$, and its Fourier transform furnishes the spectral density function:
\begin{equation}
	G(t) = - \int d\omega\, \rho(\omega) \sin\omega t
	\text.
\end{equation}
The dependence on $\beta$ (as well as the implicit dependence on $\mathcal O$) is elided, as in this work we will consider only one thermal ensemble at a time.

It is clear that $G(t)$ is of direct physical interest, as it corresponds directly to an (idealized) experiment in which a system is perturbed and then measured at a later time. The spectral density function is from a theoretical perspective the more fundamental object, and is sometimes of direct interest to experiment as well. Unfortunately, lattice Monte Carlo calculations provide neither directly. What can be measured on the lattice is the Euclidean correlator:
\begin{equation}
	G^{(E)}(t) = Z^{-1} \Tr e^{-(\beta-\tau)H} \mathcal O e^{-\tau H} \mathcal O
\end{equation}
This is of course the analytic continuation of the real-time correlator $G(t)$, with the operators being separated by an imaginary, rather than real, time interval. As a result, it is also expressible in terms of the spectral density function:
\begin{equation}
	G^{(E)}(\tau) = \int d\omega\, \rho(\omega) \frac{\cosh \omega \left(\frac \beta 2 - \tau\right)}{\sinh \frac{\beta\omega}2}
	\text.
\end{equation}

Given the mismatch between the quantity accessed by numerics ($G^{(E)}(\tau)$) and those of experimental interest ($G(t)$, $\rho(\omega)$, or smeared versions thereof), we are confronted with the task either of analytic continuation (the determination of $G(t)$) or spectral reconstruction (the determination of $\rho(\omega)$). These tasks are essentially equivalent, as given a solution to one, the other task can be accomplished via Fourier transform.

We will focus on the inverse problem---spectral reconstruction. It is often said that this problem is ill-posed (see Appendix~\ref{app:ill} for a careful discussion of the matter). This is to say that, given measurements of the Euclidean correlator at discrete time separations $G^{(E)}(\tau_i)$ with error bars $\sigma_i$, there are multiple spectral densities $\{\rho(\omega), \rho'(\omega),\ldots\}$ consistent with this numerical data. Moreover, these spectral densities can be made arbitrarily pointwise different, in the sense that $|\rho(\omega) - \rho'(\omega)|$ cannot be bounded. Therefore no finite error bars can meaningfully be provided for an estimate of $\rho(\omega)$ at any fixed $\omega$. A similar statement follows for the real-time correlator $G(t)$: pointwise estimates at non-zero time $t$ will necessarily have large errors, which are not made smaller by the provision of higher-quality Euclidean data.

It is crucial not to read the above, a ``no-go folk theorem'', too pessimistically. The obstacles mentioned above can be evaded as long as we consider only appropriately smeared integrals of the spectral density, and make use of the positivity axiom $\rho(\omega) \ge 0$. In general we will find that although $\rho(\omega)$ and $G(t)$ themselves cannot be reliably determined from Euclidean lattice data, certain integrals over the spectral density, or equivalently over the correlator, can.

A simple example at zero temperature is given by integrating the real-time correlator with an exponential decay. Define
\begin{equation}
	I_{\mathrm{exp}}(\kappa) \equiv \int_0^\infty dt\, e^{-\kappa t} G(t)
	\text,
\end{equation}
and note that the integrand vanishes as $\Re t \rightarrow \infty$ and also as $\Im t \rightarrow -\infty$. As a result we can deform the contour of integration, obtaining an alternative expression
\begin{equation}
	I_{\mathrm{exp}}(\kappa) = -i \int_0^\infty d\tau e^{i\kappa \tau} G^{(E)}(\tau)
	\text.
\end{equation}
As a result we see that the Fourier transform of the Euclidean correlator yields a set of non-trivial integrals of the real-time correlator.

By itself this is not quite sufficient to reliably evaluate $I_{\mathrm{exp}}(\kappa)$ from Euclidean data, as $G^{(E)}(\tau)$ is only available at a discrete set of Euclidean times. However, positivity of the spectral density function guarantees that $G^{(E)}(\tau)$ is monotonically decreasing for $\tau > 0$.
As a result, the estimates of the Euclidean correlator at times $\tau_i$ and $\tau_{i+1}$ respectively provide upper and lower bounds for $G^{(E)}$ in the interval $(\tau_i,\tau_{i+1})$, and the evaluation of $I_{\mathrm{exp}}(\kappa)$ is therefore fully under control.

A similar argument works at finite temperature as well, where monotonicity of the Euclidean correlator on $\tau > 0$ is replaced by piecewise monotonicity on the intervals $\tau \in (0,\frac \beta 2)$ and $\tau \in (\frac\beta 2,\beta)$. Note that we should not claim here that the bounds obtained in this way on $I_{\mathrm{exp}}(\kappa)$ are as tight as possible. Even without improving the Euclidean data, note that not only $G^{(E)}$ is monotonic, but also its derivatives---this is again a consequence of positivity of the spectral density. Tighter bounds on the integral can be obtained by making use of all consequences of the positivity constraint.

The remainder of this paper is occupied with the task of generalizing this trick to estimating more physically relevant integrals of the real-time correlator. It is perhaps not \emph{a priori} obvious which integrals are likely to be tightly bounded by given Euclidean data (together with positivity). A couple heuristic arguments, given below, serve as a useful guide.

From lattice Monte Carlo, the Euclidean correlator is given only at a finite set of Euclidean time separations\footnote{For continuous-time algorithms~\cite{Beard:1996wj,2005PhRvB..72c5122R,2011RvMP...83..349G} this is no longer strictly true, which may translate to an advantage for those algorithms with respect to the inverse problem.}, typically at even intervals. Labelling the Euclidean-time lattice spacing $\Delta \tau$, very little information regarding $\rho(\omega)$ is present for energies $\omega \gtrsim \Delta \tau^{-1}$. Asymptotically, where $\omega \gg \Delta\tau^{-1}$, only one integral of $\rho(\omega)$ is provided: the Euclidean correlator at $\tau = 0$. As a result, we should not expect to be able to bound any quantity which depends substantially on high-energy physics. In particular, the real-time correlator $G(t)$ will not be provided with any non-trivial bounds. For this reason we will consider only smeared correlators, where the smearing has been used to strongly damp all dependence on $\rho(\omega)$ at large $\omega$.

Having excluded estimates of $G(t)$ for fixed $t$ from consideration, we now turn to the notion of estimating $\rho(\omega)$ at fixed $\omega$. Again we must not expect non-trivial error bars. The reason for this is familiar from lattice spectroscopy, where we know that the spectral density is to be modelled by $\rho(\omega) = \delta(m-\omega)$ for some mass $m$, but the precise value of $m$ is unknown. Although the extracted statistical errors on the estimate of the mass may be small, the error bars on the value of $\rho(\omega)$ itself (in the region $\omega \approx m$) are clearly infinitely large. This uncertainty in the ``fine structure'' of the spectral density mostly comes from the statistical errors in the measurement of the Euclidean correlator, and is often related to the signal-to-noise problem that affects lattice spectroscopy. As in the case of the correlator, we will therefore consider only finite smearings of $\rho(\omega)$.

\section{Convex optimization of spectral bounds}\label{sec:convex}

In general we wish to bound an integral of the spectral density function of the form $\int \mathcal K(\omega) \rho(\omega) d\omega$, given lattice measurements of the Euclidean correlator. The precise choice of integration kernel $\mathcal K$ depends on the application. For this work we will consider two kernels. Most straightforwardly, a Gaussian smearing of the spectral density is defined by
\begin{equation}\label{eq:spectral-smeared}
	\begin{split}
		&\tilde \rho_\sigma(\omega) = \int_0^\infty d\omega'\, \rho(\omega') \mathcal K^\omega_\sigma(\omega')\\
		&\text{where }\mathcal K^\omega_\sigma(\omega') = \frac{1}{\sqrt{2\pi}\sigma \Phi(\frac\omega\sigma)} e^{-\frac{(\omega-\omega')^2}{2\sigma^2}}\\
		&\text{and }\Phi(x) \equiv \frac{1}{\sqrt{2\pi}} \int_{-\infty}^x dz\, e^{-\frac{z^2}{2}} = \frac {1 + \mathop{\mathrm{erf}}\big(\frac x {\sqrt{2}}\big)}{2}
		\text.
	\end{split}
\end{equation}
For directly examining real-time response, we defined a similarly smeared real-time correlator from $G(t)$ according to
\begin{equation}
	\tilde G_{\sigma}(t) \equiv \frac{1}{\sqrt{2\pi}\sigma} \int_{-\infty}^\infty dt'\,G(t)e^{-(t'-t)^2 / (2 \sigma^2)}
	\text,
\end{equation}
which may be obtained from the spectral density function according to
\begin{equation}\label{eq:G-smeared}
	\begin{split}
		\tilde G_{\sigma}(t) &= \int d\omega\,\rho(\omega) \mathcal K^t_\sigma(\omega)\\
		& \text{where } \mathcal K^t_\sigma(t,\omega) \equiv -2
		e^{-\frac{\sigma^2 \omega^2}{2}}
		\sin \omega t
	\text.
	\end{split}
\end{equation}

We will obtain bounds on such integrals of the spectral density function by writing down a \emph{convex program}: effectively, a description of the convex space of permitted spectral densities, and a linear function on the space to be extremized. This program is then solved by means of the interior-point method~\cite{boyd2004convex}, yielding an extremal point and therefore a bound on either $\tilde \rho_\sigma(\omega)$ or $\tilde G_\sigma(t)$.

The specifics of the convex programs to be constructed are detailed below. However, in general, there are two sorts of constraints. One set comes from the requirement that $\rho(\omega)$ be everywhere non-negative; the other comes from the knowledge of the Euclidean lattice data. The Euclidean lattice data provides a (noisy) measurement of integrals of the spectral density function of the form
\begin{equation}
	\begin{split}
		G^{(E)}(\tau) &= \int_0^\infty d\omega\,\rho(\omega) K(\tau,\omega)\\
		&\text{where } K(\tau,\omega) \equiv\frac{\cosh \omega \left(\frac \beta 2 - \tau\right)}{\sinh \frac{\beta\omega}2}
		\text.
	\end{split}
\end{equation}
At first, we will treat this Euclidean-time data as exact, resulting in a set of linear constraints on the spectral density function. The realistic case, in which the Euclidean-time measurements are noisy, is treated in the final subsection.

\subsection{Inversion of exact data}
Let us begin by considering the case where the exact Euclidean correlator, free of statistical errors, is known at a finite set of imaginary time separations $\tau_i$. Without use of the fact that $\rho(\omega) \ge 0$, the integral of $\rho$ with $\mathcal K$ cannot be bounded due to the discretization of imaginary time.

If the correlator is known at $N$ distinct imaginary times $\tau_1,\ldots,\tau_N$, then the spectral density function is constrained by $N$ linear constraints of the form
\begin{equation}
	\begin{split}
		\int_0^\infty d\omega\, \rho(\omega) K_i(\omega) &= C_i\\
		\text{where }&K_i(\omega) \equiv \frac{\cosh \omega \left(\frac \beta 2 - \tau_i\right)}{\sinh \frac{\beta \omega}{2}}
		\text.
	\end{split}
\end{equation}
The task of determining $\int \rho(\omega) \mathcal K(\omega)$ for general $\mathcal K(\cdot)$ is ill-posed precisely because $\mathcal K$ may be linearly independent of $\{K_i\}$. Including the positivity constraint reduces the uncertainty in the integral to a finite range, which shrinks with improved Euclidean data.

Including positivity, we may express the optimization of a hypothetical integral as a convex program:
\begin{equation}\label{eq:spectral-lp}
	\begin{split}
		\text{minimize }&\int_0^\infty d\omega\, \rho(\omega) \mathcal K(\omega)\\
		\text{subject to }&\rho(\omega) \ge 0\\
		\text{and }&\int_0^\infty d\omega\, \rho(\omega) K_i(\omega) = C_i
		\text.
	\end{split}
\end{equation}
Taking as a given that this optimization problem may be solved efficiently, we may bound (for example) the smeared real-time correlator $\tilde G_\sigma(t)$ straightforwardly. The optimization problem (\ref{eq:spectral-lp}) is solved twice: once for $\mathcal K(\omega) = \mathcal K^t_\sigma(\omega)$ as defined in (\ref{eq:G-smeared}) to obtain a lower bound, and once for $\mathcal K(\omega) = -\mathcal K^t_\sigma(\omega)$ to obtain an upper bound.

The optimization problem (\ref{eq:spectral-lp}) is an infinite-dimensional variant of a \emph{linear program}. A general linear program\footnote{The earliest unambiguous appearance of linear programs was in the Soviet Union, motivated by the problem of central planning~\cite{kantorovich1960mathematical}; the same idea appeared a few years later in the West, due to Dantzig~\cite{dantzig1983reminiscences}, and soon became central to operations research.} may be written in the form:
\begin{equation}\label{eq:augmented-lp}
	\begin{split}
		\text{minimize }&c^T x\\
		\text{subject to }&y_i^T x = b_i\\
		\text{and }&x \ge 0\text.
	\end{split}
\end{equation}

A practical algorithm for solving linear programs is given by the simplex method~\cite{dantzig1951maximization}. It should go without saying that the simplex method expects to operate on a \emph{finite} linear program; the optimization problem (\ref{eq:spectral-lp}) involves infinitely many degrees of freedom. In practice one need only truncate the linear program above by considering the value of $\rho(\omega)$ at some finite set of energies, and replacing the integrals by sums. As algorithms for solving linear programs are quite fast (easily scaling above $10^4$ degrees of freedom), the discretization error is negligible.

One may nonetheless worry that solving the discretized linear program does not amount to a proof; in other words, that there may be some particularly diabolical $\rho(\omega)$, not well-approximated by our discretization, that yields a substantially different integral $\int \rho(\omega) \mathcal K(\omega)$. In fact it is possible to prove ``by hand'' that this is not the case, and the discretization described above converges rapidly. However, by making use of Lagrange duality, the infinite-dimensional linear program above can be converted to a finite-dimensional convex optimization problem, with no discretization or approximation needed. Solving the resulting ``dual problem'' is at once aesthetically pleasing and far more computationally efficient.

The next subsection introduces Lagrange duality in the general context, and subsequently we will formulate a convex program that takes into account statistical errors and derive its Lagrange dual.

\subsection{Lagrange duality}
Consider a typical convex optimization problem\footnote{This subsection is only a brief introduction to the topic; a standard, and far more thorough, reference is~\cite{boyd2004convex}.}, which we will write in the form:
\begin{equation}\label{eq:convex-problem}
	\begin{split}
		\text{minimize }&c^T x\\
		\text{subject to }&h_i(x) \ge 0\text.
	\end{split}
\end{equation}
Here the functions $h_i$ are required to each be convex, rendering the region of $x$ consistent with the constraints (termed the \emph{feasible region}) itself a convex region. The objective function is assumed to be linear for convenience.

Suppose $x$ is an $N$-dimensional vector, and let there be $M$ inequality constraints defined by $M$ functions $h_i$. We will convert this minimization problem over (a subset of) $\mathbb R^N$ to a maximization problem over $\mathbb R^M$. This is likely to be computationally advantageous when $N \gg M$. We begin by defining the \emph{Lagrange function}:
\begin{equation}
	L(x,\lambda) = c^T x - \sum_i \lambda_i h_i(x)
	\text.
\end{equation}
Above, the components of the vector $\lambda$ are termed \emph{Lagrange multipliers}. Note that, for fixed $x$, we have
\begin{equation}
	\max_{\lambda \ge 0} L(x,\lambda) = \begin{cases}
		c^T x & h_i(x) \ge 0 \text{ for all } i\\
		\infty & \text{otherwise.}
	\end{cases}
\end{equation}
Therefore the original optimization problem can be written as a minimization (over $x$) of this maximization (over $\lambda \ge 0$): the minimization will always avoid the region where the constraints on $x$ are violated. We write the solution to (\ref{eq:convex-problem})---henceforth referred to as the \emph{primal problem}---as
\begin{equation}
	p^* = \min_x \max_{\lambda \ge 0} L(x,\lambda)\text.
\end{equation}
It is straightforward to prove that for any function $f(x,y)$, we have an inequality
\begin{equation}\label{eq:weak-duality}
	\min_x \max_y f(x,y) \ge \max_y \min_x f(x,y) \text.
\end{equation}
This theorem motivates the definition of a dual optimum, defined by switching the order of the minimization and maximization above:
\begin{equation}
	d^* = \max_{\lambda \ge 0} \min_x L(x,\lambda)\text.
\end{equation}
When provided concrete forms for the inequality constraints $h_i(x)$, and therefore for the Lagrange function, it is often possible to evaluate the inner minimization in closed form. This defines a \emph{Lagrange dual function} $g(\lambda) = \min_x L(x,\lambda)$, which is in turn used to define the \emph{dual problem}:
\begin{equation}
	\begin{split}
		\text{maximize }&g(\lambda)\\
		\text{subject to }&\lambda_i \ge 0
		\text.
	\end{split}
\end{equation}
Eq.~(\ref{eq:weak-duality}) establishes that the solution to the dual is a lower bound on the solution to the primal. This statement is referred to as \emph{weak duality}. Recall that a point that satisfies the constraints of any given optimization is referred to as a feasible point. Denote the dual optimum point $\lambda^*$, and the primal optimum point $x^*$. Since for any dual-feasible $\lambda$ we have $g(\lambda) \le \lambda^*$, and for any primal-feasible $x$ we similarly have $c^T x \ge c^T x^*$, it follows that $g(\lambda) \le c^T x$ as long as both $\lambda$ and $x$ are feasible. From a practical perspective this is extraordinarily convenient: a dual-feasible point $\lambda$ provides a lower bound on the primal optimum $p^*$, which can be efficiently verified without the need to perform any optimization whatsoever.

Weak duality holds for any primal/dual pair constructed as above. Often, \emph{strong duality} holds as well, meaning that the inequality is saturated at $p^* = d^*$. Fairly general sufficient conditions are established by Sion~\cite{pjm/1103040253}. A stronger sufficient condition, often satisfied in practice, was formulated by Slater~\cite{RePEc:cwl:cwldpp:80}: if the primal problem has a strictly feasible point (that is, a point at which all inequalities are satisfied as strict inequalities), then the primal optimum and dual optimum are equal. The optimization problems described in this work all satisfy this condition. The practical consequence of strong duality is that no information is lost in passing from the primal to the dual.

Following the procedure above, the Lagrange dual of (\ref{eq:spectral-lp}) may be readily derived, yielding the following convex program:
\begin{equation}\label{eq:spectral-lp-dual}
	\begin{split}
		\text{maximize }& \ell_i C_i\\
		\text{subject to }& \lambda(\omega) \equiv \mathcal K(\omega) - \sum_i \ell_i K_i(\omega) \ge 0
		\text.
	\end{split}
\end{equation}
In this case the Lagrange dual has a fairly intuitive interpretation, and can be understood without reference to the primal program or any of the machinery of convex optimization. Using the above definition of $\lambda$, and the fact that both $\rho(\omega)$ and $\lambda(\omega)$ are non-negative, we have:
\begin{equation}
	0 \le \int \rho \lambda = \int_0^\infty d\omega\,\rho(\omega) \mathcal K(\omega) - \ell_i C_i\text.
\end{equation}
It immediately follows that $\ell_i C_i$ is a lower bound on the integral of interest, $\int \rho \mathcal K$.

\subsection{Frequentist inversion}\label{ssec:frequentist}

In the presence of statistical errors, the measurements of the Euclidean correlator no longer provide linear equality constraints, and the optimization problem described previously is no longer directly applicable\footnote{An attentive reader might object that there is an obvious extension of the above algorithm: compute resampled means of Euclidean correlators \emph{\`a la} statistical bootstrap, and apply the above procedure for analytic continuation of exact data to each resampled mean. Unfortunately, numerical experiments reveal that a typical sample of the Euclidean correlator is not produced by any non-negative $\rho(\omega)$.}. This subsection describes the necessary modifications to enable the bounding algorithm to operate on Monte Carlo-generated data. As before, we will construct a convex program to represent the space of permitted spectral densities, but because the error bars on the Euclidean data describe a region rather than a hypersurface, the program will not be a linear program but instead a more general convex program.

The positivity constraint is the same as always: $\rho(\omega) \ge 0$. Intuitively, the constraint from the Euclidean data is now that the distance between the measured correlator and the correlator computed from the spectral density, described by the vector
\begin{equation}
	v_i[\rho] = \int K_i(\omega) \rho(\omega) \,d\omega - C_i\text,
\end{equation}
is not too large relative to the statistical errors in the estimate of $C$. To operationalize the notion of ``not too large'' we must pick a single statistic---a function of $\rho$ on which we will place an upper cutoff. As the errors on $C$ are approximately Gaussian, it is natural to denote the covariance matrix $\Sigma$ and define the statistic
\begin{equation}\label{eq:statistic}
	\begin{split}
		F[\rho] &= v[\rho]^T M v[\rho]\\
		&\text{where } M \equiv \left[\Sigma + \epsilon ||\Sigma|| \right]^{-1}
	\end{split}
\end{equation}
The regulator $\epsilon$, inserted for numerical stability, is taken to be $10^{-6}$ throughout.

By resampling the Monte Carlo data we can select a maximum value $F_{\mathrm{max}}$ for $F[\rho]$, corresponding to a $99\%$ confidence region. To expand on this point, let $C_i$ be the estimates of the Euclidean correlator obtained by averaging all Monte Carlo data, and let $C^{(k)}_i$ be $K$ different estimates of the Euclidean correlator obtained by averaging resampled data. We may then compute $K$ different error statistics $F_k$, defined by
\begin{equation}
	\begin{split}
		F_k &= (v^{(k)})^T M v^{(k)}
		\text{ where } v^{(k)}_i = C^{(k)}_i - C_i
		\text.
	\end{split}
\end{equation}
This gives us empirical access to the distribution of values $F[\rho^*]$ takes, where $\rho^*$ is the true spectral density. We take $K=1000$ samples, sort in decreasing order, and define $F_{\mathrm{max}}$ to be the tenth sample. Now the condition $F[\rho] \le F_{\mathrm{max}}$ may be thought of as a $99\%$ confidence interval.

Putting this all together, we arrive at the following (primal) convex program for a lower bound on an arbitrary integral of the spectral density:
\begin{equation}\label{eq:spectral-primal}
	\begin{split}
		\text{minimize }& \int_0^\infty d\omega\, \mathcal K(\omega) \rho(\omega)\\
		\text{subject to }& \rho(\omega) \ge 0 \\
		\text{and }& F[\rho] \le F_{\mathrm{max}}
		\text.
	\end{split}
\end{equation}
Solving this program yields a $99\%$ confidence interval\footnote{Strictly speaking, the interval may be larger, but no smaller, than a true $99\%$ confidence interval.} for the desired integral. Note that the error bars obtained this way include both statistical and systematic errors, with the former coming from the Monte Carlo sampling and the latter from the ill-posed inversion. The systematic errors cannot be approximated by a Gaussian. (This is why we have chosen to present a $99\%$ confidence interval, instead of a more traditional $1\sigma$ interval---the latter would provide no useful information about the former.)

With an infinite number of degrees of freedom to be optimized over, this form of the convex program is not computationally convenient. Instead, we will work with the Lagrange dual:
\begin{equation}
\label{eq:spectral-dual}
	\begin{split}
		\text{maximize }&\ell^T C - \frac{F_{\mathrm{max}}}{4\mu} \ell^T M^{-1} \ell - \mu\\
		\text{subject to }& \mathcal K(\omega) - \sum_i \ell_i K_i(\omega) \ge 0\\
		\text{and }&\mu \ge 0
	\text.
	\end{split}
\end{equation}
Here the Lagrange multipliers are a vector $\ell \in \mathbb R^N$ (where there are $N$ Euclidean time separations measured) and $\mu \in \mathbb R$. A derivation of this dual program is provided in Appendix~\ref{app:dual}. It has two important properties, mentioned in the preceding subsection. First, any feasible point of the dual provides a lower bound on the optimum, so that it is easy to verify that obtained bounds are valid. Second, the two optima are equal, meaning that bounds provable via the dual can be made arbitrarily tight. Therefore, no information is lost in solving the dual in place of the primal.

The dual program (\ref{eq:spectral-dual}) is an optimization program over a finite dimensional space, but still with an infinite number of constraints, as the function $\mathcal K - \ell \cdot K$ is required to be non-negative at every $\omega \in [0,\infty)$. Due to the restricted form of this function, these constraints can be checked in finite time\footnote{It may appear surprising that an infinite set of constraints can be checked in finite time, but the principle is already familiar. Consider the following infinite set of constraints on a $N \times N$ matrix $A$: $\forall v \in \mathbb R^N$, $v^T A v \ge 0$. These constraints can be checked in finite (in fact polynomial) time by diagonalizing $A$.}; the procedure for performing the optimization is detailed in Appendix~\ref{app:ipm}.

\section{Smeared spectral densities}\label{sec:density}
\begin{figure*}
	\centering
	\includegraphics[width=0.48\linewidth]{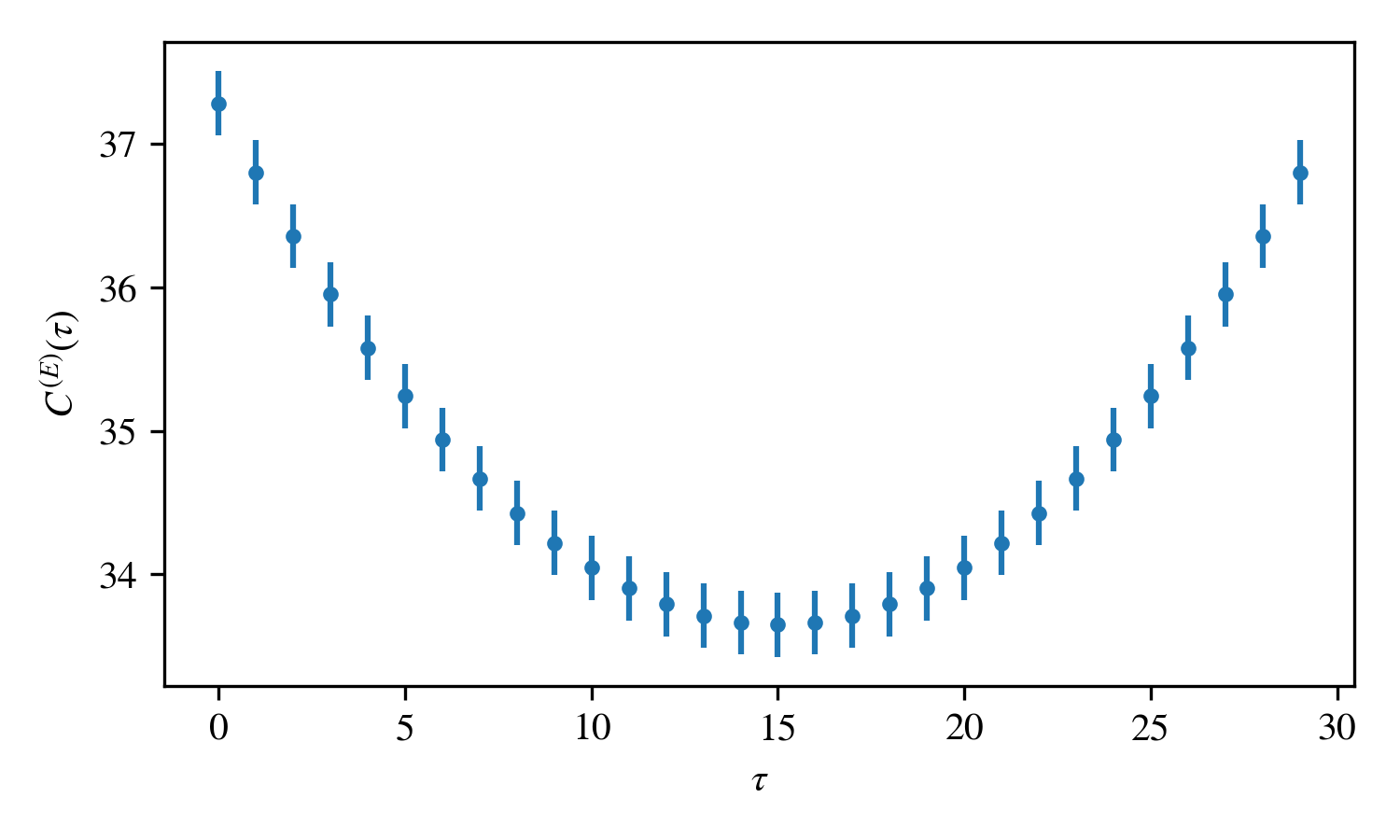}
	\hfill
	\includegraphics[width=0.48\linewidth]{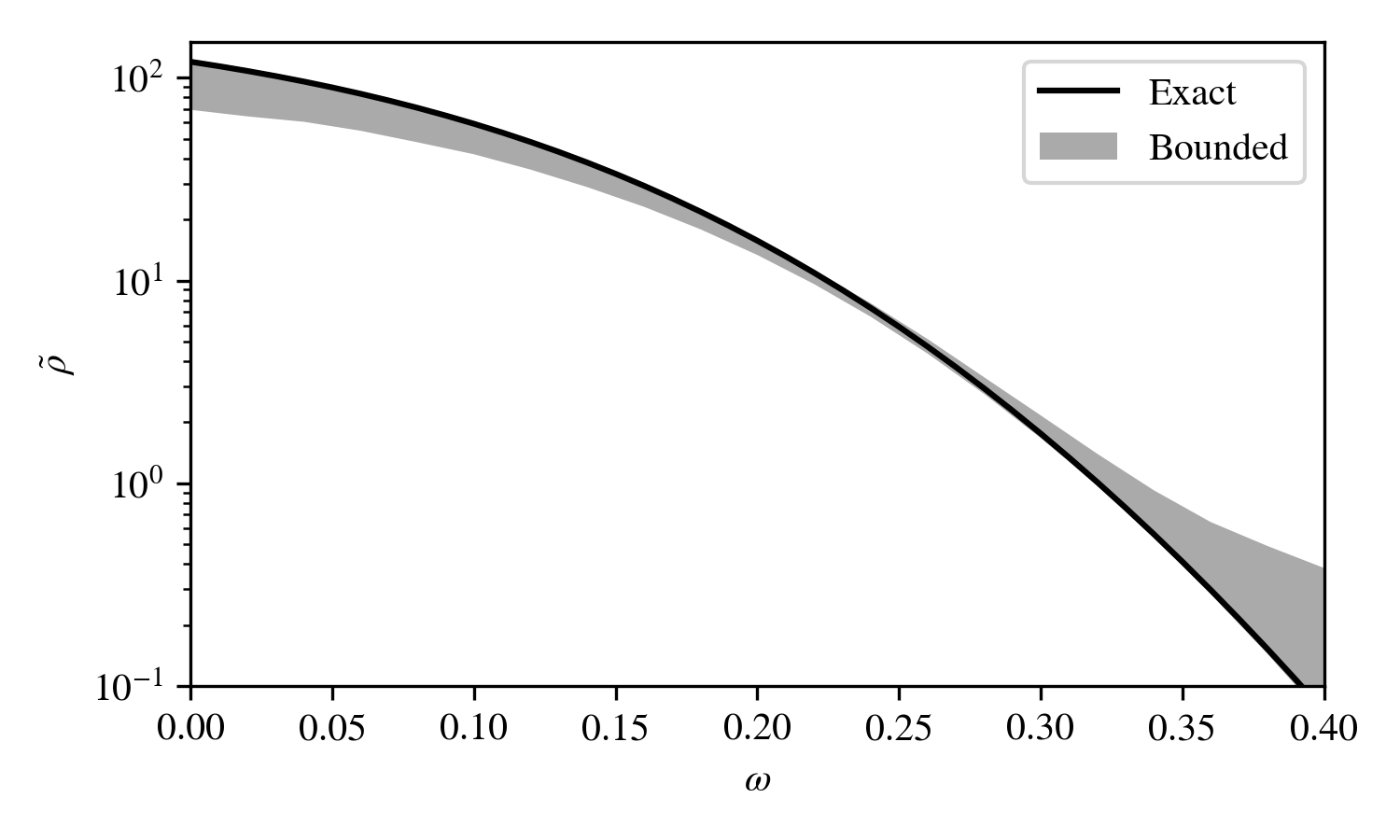}
	\caption{Deriving bounds on the (smeared) spectral density function of the anharmonic oscillator, from the Euclidean correlator. At left, the Euclidean correlator of the lattice theory defined by Eq.~(\ref{eq:action}), with $\omega^2 = 10^{-4}$, $\lambda = 10^{-5}$, and an inverse temperature of $\beta = 30$. The correlator is measured on $3 \times 10^4$ decorrelated samples. At right, the exact and estimated smeared spectral densities. The smearing is defined by a Gaussian with width of $\sigma = 0.1$.\label{fig:spec}}
\end{figure*}

For demonstration and testing we will study the quantum anharmonic oscillator, for which high-precision (effectively exact) numerical results may be obtained by straightforward diagonalization of the Hamiltonian. We will reproduce a smeared spectral density in this section, and a smeared real-time correlation function in the next, in each case comparing the result obtained by diagonalization to the bounds given by constrained inversion as described in the previous section.

The Hamiltonian of our system is
\begin{equation}\label{eq:hamiltonian}
	\hat H_{\mathrm{osc}} = \frac 1 2 \hat p^2 + \frac{\omega^2}{2} \hat x^2 + \frac\lambda 4 \hat x^4
	\text.
\end{equation}
where $\omega$ sets the gap of the system in the absence of the deformation $\lambda$.

Performing the usual Euclidean-time Trotter-Suzuki procedure yields a Euclidean lattice action:
\begin{equation}\label{eq:action}
	S_{\mathrm{osc}}(x) = \sum_{\langle r r'\rangle} \frac{\left(x_r - x_r'\right)^2}{2} + \sum_r \frac{\omega^2}{2} x_r^2 + \frac{\lambda}{4} x_r^4
	\text.
\end{equation}
The first summation is performed over all (unordered) pairs of adjacent sites, and the second over all sites. All parameters are implicitly in units of the lattice spacing, and the path integral with respect to the action Eq.~(\ref{eq:action}) is a good approximation in the simultaneous limits $m,\lambda^{1/3} \ll 1$.

Sampling a lattice of $\beta$ sites yields expectation values evaluated in the quantum thermal ensemble defined by the density matrix $e^{-\beta \hat H}$. To estimate the spectral density function we will use only the Euclidean two-point function of the operator $\hat x$, evaluated on the lattice by
\begin{equation}
	C^{(E)}_{\mathrm{osc}}(\tau) = \langle x(\tau) x(0) \rangle\text.
\end{equation}
In principle many other correlators are available and might be used to further constrain the spectral density; this possibility is discussed at slightly more length in Section~\ref{sec:discussion}.

For this demonstration, we work with Hamiltonian parameters of $\omega^2 = 10^{-4}$ and $\lambda = 10^{-5}$, at an inverse temperature of $\beta = 30$. The gap between the ground state and the first excited state is $\Delta E \approx 0.02498$ in lattice units.

This Euclidean correlator measured on $3 \times 10^4$ (decorrelated) samples is shown in the left panel of Figure~\ref{fig:spec}. Note that not all information used by the inversion procedure is present in that figure: the displayed statistical errors correspond only to the diagonal part of the covariance matrix, while the full covariance matrix is used to compute the inversion.

The right panel of Figure~\ref{fig:spec} shows the result of the inversion. Let us briefly review the procedure for completeness. First, the covariance matrix of the Euclidean correlator is determined, and a statistic $F$ defined according to Eq.~(\ref{eq:statistic}). The cutoff on the statistic is selected (by statistical bootstrap) to define a $99\%$ confidence interval. For each value of $\omega$ plotted, two convex programs of the form (\ref{eq:spectral-dual}) are solved using the interior-point method described in Appendix~\ref{app:ipm}, corresponding to the upper and lower bounds.

Only for a finite number of values of $\omega$ are bounds computed in this manner. For display purposes, linear interpolation is used in producing the right panel of Figure~\ref{fig:spec}. In practice the bounds obtained are sufficiently smooth in $\omega$ that the shaded region may be taken as the ``allowed'' region.

This is compared against a nearly exact smeared spectral density computed from the diagonalization of the Hamiltonian of Eq.~(\ref{eq:hamiltonian}), in the basis formed by the $300$ lowest eigenstates of the corresponding harmonic oscillator ($\lambda = 0$). The smeared spectral density is computed from
\begin{widetext}
\begin{equation}
	\tilde\rho_\sigma(\omega) = \frac 1 {\sqrt{2\pi}\sigma \Phi\big(\frac\omega\sigma\big)}\int_0^\infty d\omega'\, \rho(\omega') e^{-\frac{(\omega-\omega')^2}{2\sigma^2}}
		\text{ where } \rho(\omega) = 2 Z^{-1} \sinh \frac{\beta\omega}{2} \sum_{i < j} |\mathcal O_{ij}|^2 e^{-\beta \frac{E_i + E_j}{2}} \delta(E_j-E_i-\omega)
	\text.
\end{equation}
\end{widetext}
Above $Z$ refers to the partition function, the sum is taken over all (ordered) pairs of eigenstates, and the modified error function $\Phi$ is defined as in Eq.~(\ref{eq:spectral-smeared}).

Because the underlying inversion problem remains ill-posed, no part of this procedure provides a point estimate of the (smeared) spectral density. As long as the exact value lies within the bound, one can make no statements regarding the quality of agreement. The exact value is expected to lie outside of the bound $<1\%$ of the time.

As mentioned already, bounds of this form on an integral $\int \mathcal K \rho$ are information-theoretically complete, in the sense that for any real number $I'$ lying within the interval defined by the bounds, a non-negative spectral density can be found which reproduces both the Euclidean correlator and the integral $I\prime$. This does not mean that the shaded region in Figure~\ref{fig:spec} is information-theoretically complete. Curves may be drawn which lie entirely within the shaded region, and yet are not consistent with the Euclidean correlator.

As a practical consequence of the above consideration, the real-time correlator (or any other integral of the spectral density) is not to be estimated from the data in the right panel of Figure~\ref{fig:spec}. Each integral must be bounded separately, by a new convex program. We turn to this task for the real-time correlator in the next section.

\section{Real-time dynamics}\label{sec:realtime}
\begin{figure*}
	\centering
	\includegraphics[width=0.48\linewidth]{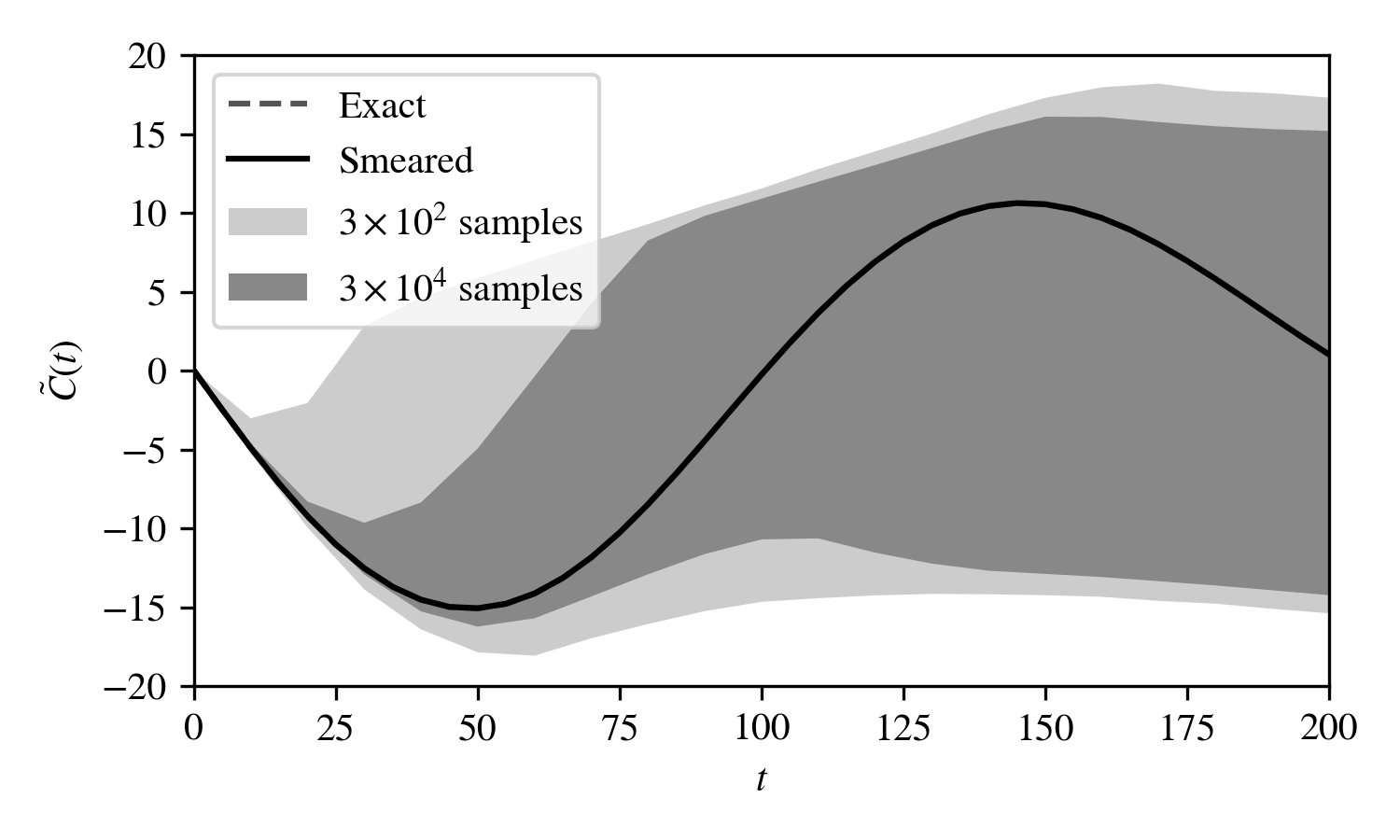}
	\hfill
	\includegraphics[width=0.48\linewidth]{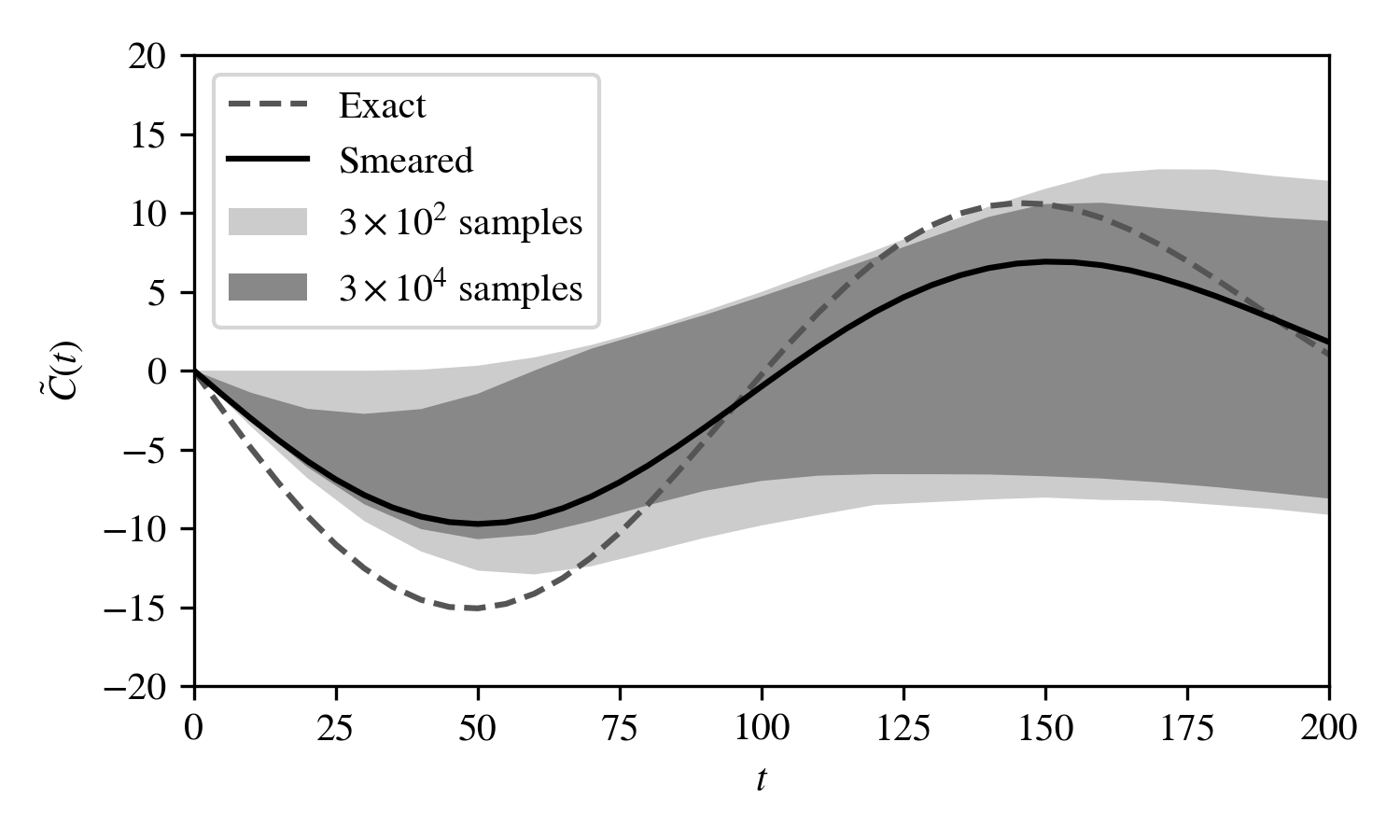}
	\caption{Determination of the smeared real-time correlator $\tilde G(t)$ from Euclidean data. The Euclidean data used is the same as for Figure~\ref{fig:spec}. At left, the correlator is smeared with a standard deviation of $\sigma=1$---there is no visible distinction between the exact and smeared correlators. Two sets of bounds on the correlator are shown; one using the full $3 \times 10^4$ samples, and the other using only $3 \times 10^2$ decorrelated samples. At right is the same plot, but for a smearing set by $\sigma=30$.\label{fig:rt}}
\end{figure*}

In this section we continue to study the Hamiltonian of Eq.~(\ref{eq:hamiltonian}), with the same lattice parameters. Rather than determine the smeared spectral density, we will estimate the smeared real-time correlator, as defined by Eq.~(\ref{eq:G-smeared}).

Similar to the case of the spectral density function, the real-time correlator
\begin{equation}
	C_{\mathrm{osc}}(t) = \langle \hat x(t) \hat x(0) \rangle
\end{equation}
may be computed to high (effectively exact) precision by diagonalizing a truncated Hamiltonian. In Figure~\ref{fig:rt} such a calculation is compared against the inversion procedure described above, for four different combinations of hyperparameters. For each time $t$, the shaded region shows the interval of achievable smeared correlation functions $\tilde G(t)$. For any point in the shaded interval, one may find a spectral density, non-negative and consistent with the Euclidean data, whose Fourier transform yields a real-time correlator that, when smeared, goes through that point.

The error bars obtained on the smeared real-time correlator, particularly at late times, are far larger than those that were obtained on the smeared spectral density. After a time of $t \sim 2 \pi m^{-1}$, the bounds obtained are completely uninformative, indicating that there is not enough information in the Euclidean correlator to provide any meaningful constraints at such times. This inferior performance is due to the rapidly oscillating integration kernel, which introduces sensitive dependence of $\tilde G(t)$ on ``fine structure'' of the spectral density. The Euclidean correlator does not share this sensitive dependence, and therefore it does not contain enough information to provide tight constraints at large times.

There is a connection here to the signal-to-noise problem of hadronic spectroscopy. The value of a mass depends on the precise location of a peak in the spectral density (rather than just the integral over some region). If we wish to determine some low-lying mass, the most sensitive part of the Euclidean correlator is that measured at large time separations, and so the late-time noise in the correlator is the chief obstacle. The same is apparently true in determining the real-time correlator.

The most obvious trend is that the systematic errors become dramatically larger at later times, eventually saturating. This is alleviated somewhat by using higher statistics, although it is clear that even with $\sim 10^6$ samples there will not be enough information in the Euclidean correlator to credibly resolve a full period of oscillation.

Increasing the amount of smearing above $\sigma\sim 1$ does not shrink the error bars, at least relative to the amplitude of the oscillation. In fact, at early times $t \lesssim 10$ the errors are made markedly larger when $\sigma$ is larger. This may be explained heuristically by noting that the smeared correlator $\tilde C_\sigma(t)$ depends most strongly on the true real-time correlator at times ranging over the interval $[t-\sigma,t+\sigma]$. Where $\sigma \gg t$, this introduces dependence on much later times, where less information is available.

Typically, approximation-free classical computational methods for accessing real-time quantum dynamics have a scaling that is exponential in the lattice volume of the system being considered. This is most apparent with methods based on diagonalizing a truncated Hamiltonian (such as the one used to produce the comparison in Figure~\ref{fig:rt}). Path integral-based methods similarly encounter a sign problem which is exponential in the spacetime volume, although the extra exponential scaling in time is compensated for by the improvements that can be made to the average sign through contour deformations.

This exponential scaling in lattice volume is apparently absent\footnote{The possibility remains that this method has exponential scaling in physical volume.} for the method described here. The computational resources required to obtain a Euclidean correlator at usable precision do not drastically increase, for example, when going from $0+1$-dimensional quantum mechanics to a $1+1$-dimensional scalar field theory. To demonstrate this concretely, consider the case of $\phi^4$ theory in $2$ spacetime dimensions. The lattice action is
\begin{equation}\label{eq:scalar-action}
	S_{\mathrm{scalar}}(\phi) = \sum_{\langle r r'\rangle} \frac{\left(\phi_r - \phi_r'\right)^2}{2} + \sum_r \frac{m^2}{2} \left[\phi_r^2 + \frac{\lambda}{4} \phi_r^4\right]
	\text.
\end{equation}
As before, the first summation is over all pairs of neighboring sites on a two-dimensional lattice.

This system is of course closely analogous to the anharmonic oscillator considered above. We will consider the two-point function and associated spectral density of the averaged field operator
\begin{equation}
	\bar\phi(\tau) = \sum_x \phi_{x,\tau}
	\text.
\end{equation}
To describe a moderately high-temperature system, we work on a lattice with $L=100$ sites in the spatial direction and $\beta = 30$ sites in the temporal direction. The lattice parameters are $m^2 = 10^{-4}$ and $\lambda = 10^{-5}$. The data used for the inversion are $10^4$ decorrelated samples obtained via a Wolff-type embedded cluster algorithm~\cite{Wolff:1988uh}.

\begin{figure}
	\centering
	\includegraphics[width=0.95\linewidth]{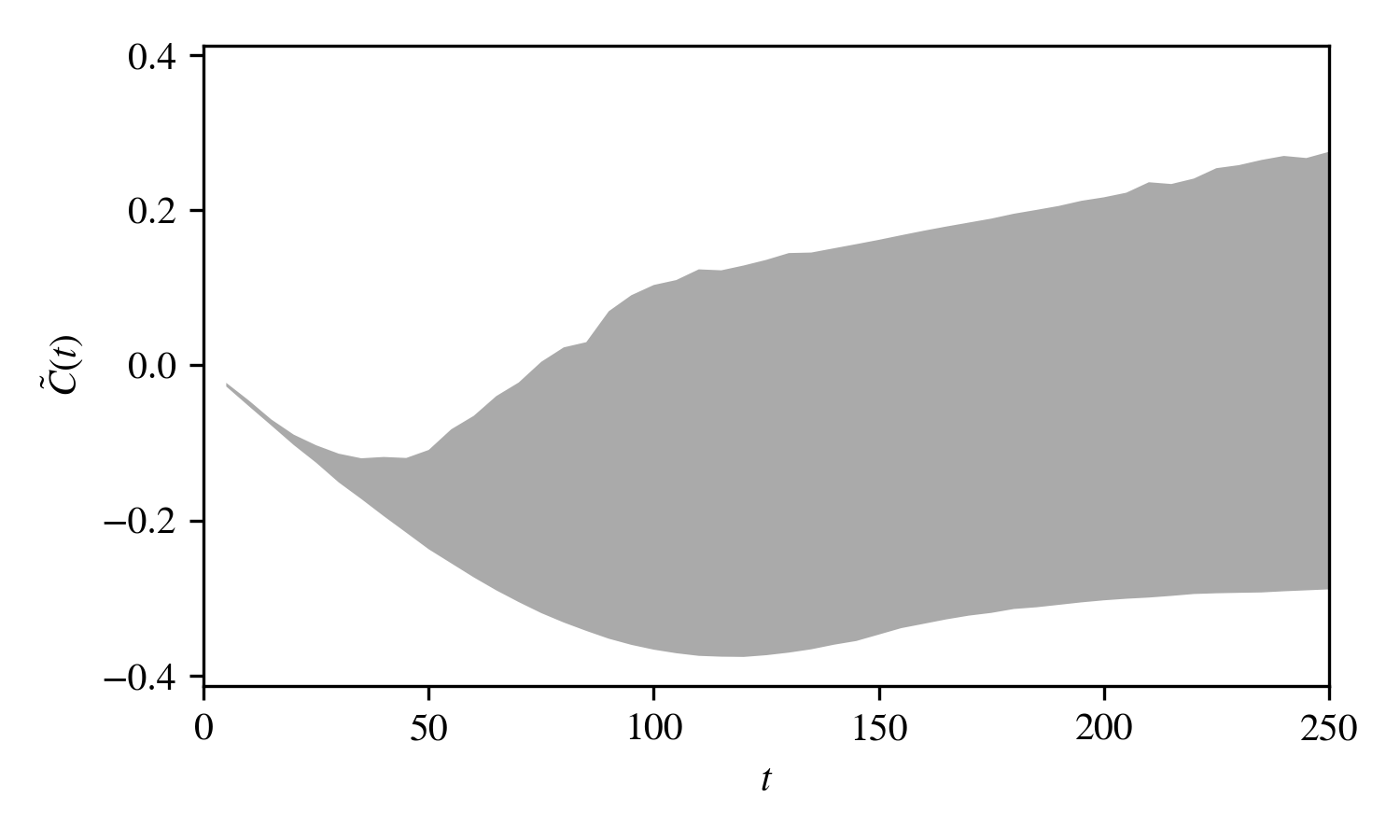}
	\caption{Bounds on a real-time correlator in $1+1$-dimensional scalar field theory, obtained through the convex program described in the text. A total of $10^4$ decorrelated samples on a $100 \times 30$ lattice are used.\label{fig:scalar}}
\end{figure}

The real-time correlator under consideration is $G(t) = \Im \langle \bar\phi(t) \bar\phi(0)\rangle$. The bounds on the corresponding smeared real-time correlator, defined according to Eq.~(\ref{eq:G-smeared}) with a standard deviation of $\sigma = 5$, are shown in Figure~\ref{fig:scalar}. As in the case of the $0+1$-dimensional, quantum-mechanical system, with these statistics the confidence interval rapidly becomes large enough to yield no useful information. What is remarkable is that the performance is not markedly worse, despite a lattice volume two orders of magnitude larger.

\section{Low temperatures}\label{sec:cold}
\begin{figure*}
	\centering
	\includegraphics[width=0.48\linewidth]{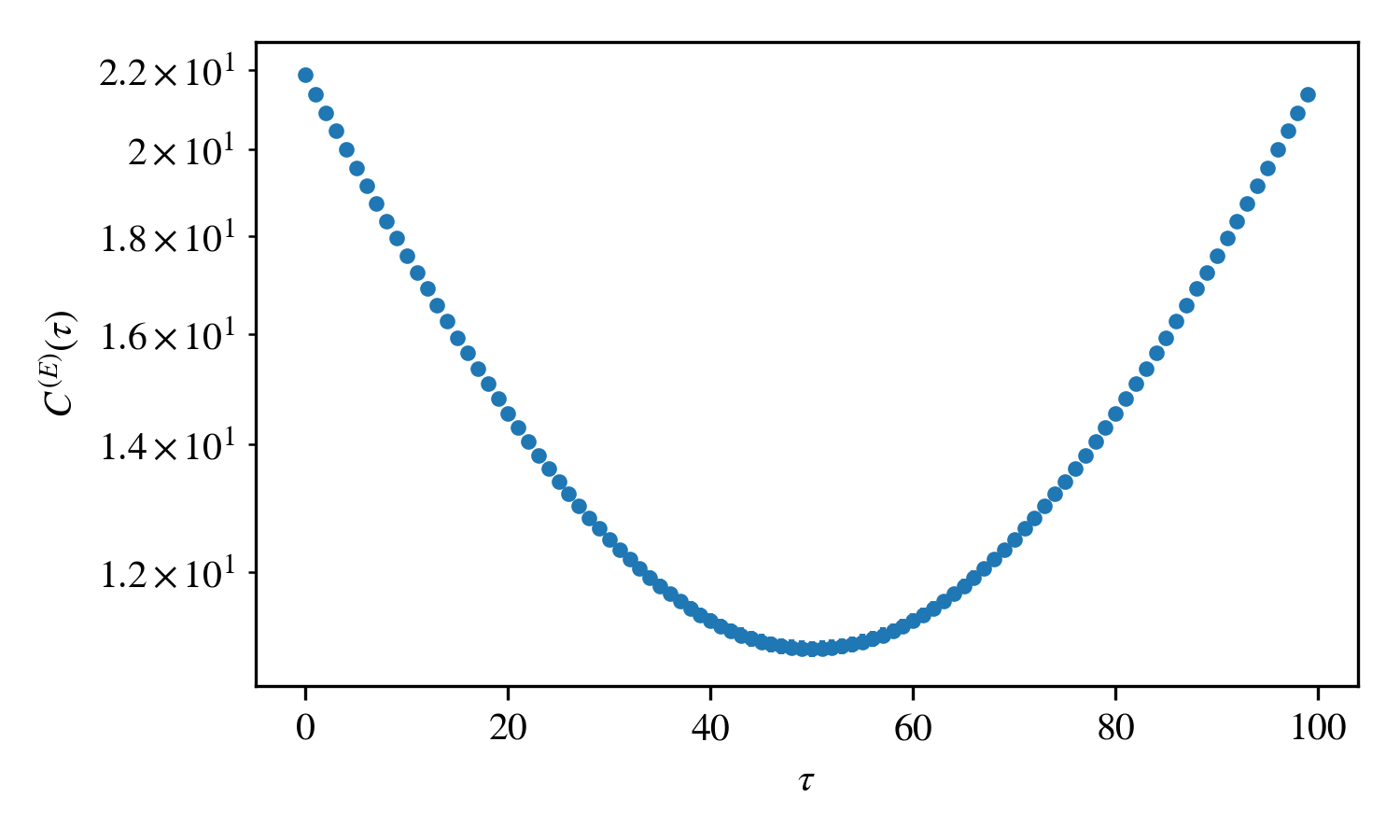}
	\hfill
	\includegraphics[width=0.48\linewidth]{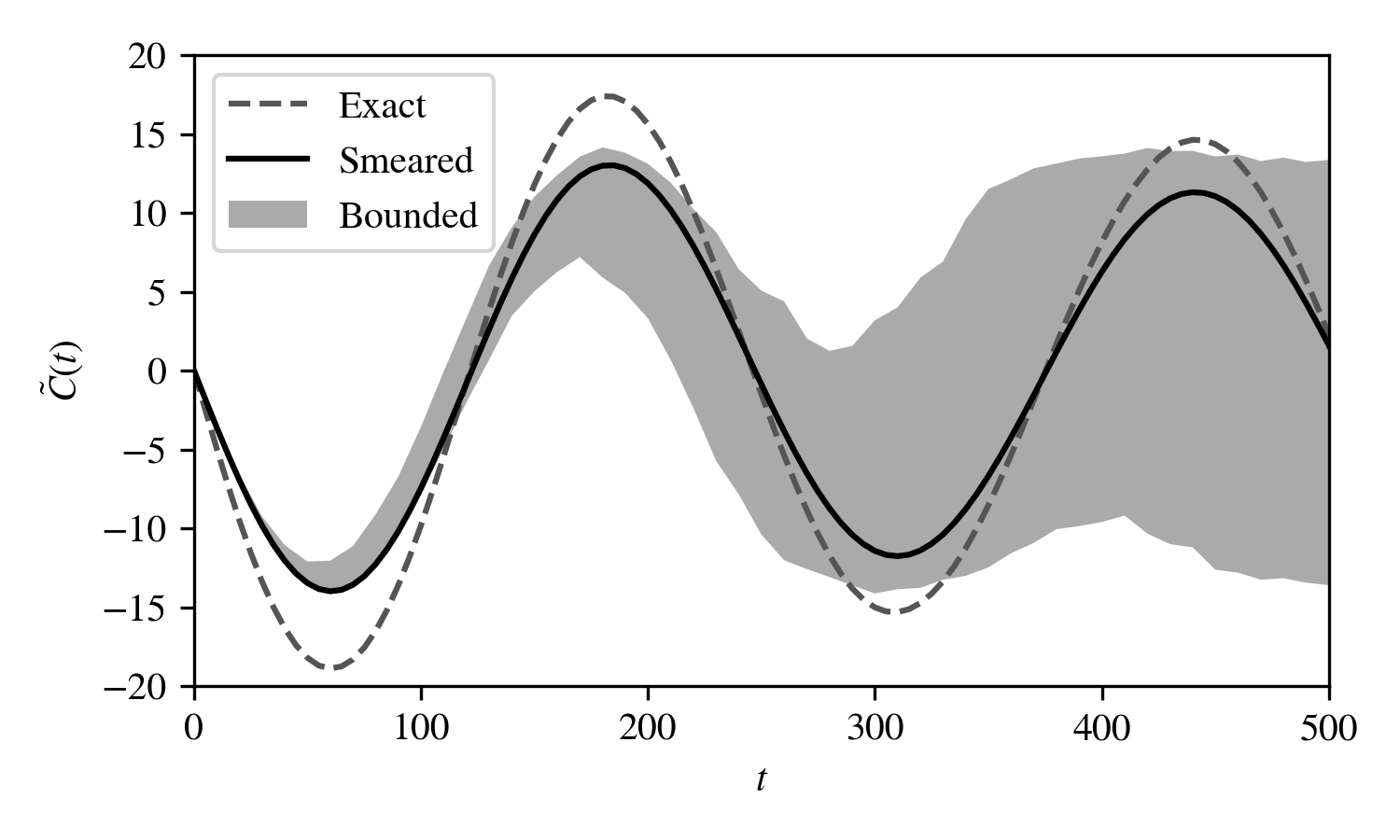}
	\caption{Determination of a smeared real-time correlator of the anharmonic oscillator at low temperature. The Hamiltonian parameters are the same as before---$\omega^2=10^{-4}$ and $\lambda = 10^{-5}$---but the inverse temperature is now $\beta=100$. Substantially higher precision is obtained in the Euclidean correlator (left panel---error bars plotted but not visible at this resolution), leading to substantially higher precision in the estimate of the real-time correlator (right panel). The smearing is defined with a standard deviation of $30$; a total of $3 \times 10^4$ decorrelated samples are used.\label{fig:cold}}
\end{figure*}
The calculations in the previous two sections were deliberately performed at high temperatures, with $\beta \Delta E < 1$. This results in nontrivial spectral densities and interesting physics. It is, after all, at high temperatures where a hydrodynamic theory of quantum matter becomes applicable at even over short distances---determination of hydrodynamic coefficients was the motivating example for this work.

Also at high temperatures, hyperbolic fits to the Euclidean correlator are particularly noisy. As the temperature is decreased, the Euclidean correlator is observed at larger imaginary time separations where excited states are suppressed. For sufficiently low temperatures, only one excited state contributes (in a gapped theory), and its mass and overlap can be fit with high precision. It is therefore unsurprising that the real-time correlator is also determined with higher precision and at later times, when the temperature is lower.

Figure~\ref{fig:cold} demonstrates this in the anharmonic oscillator. The same Hamiltonian parameters are used as above, but with a substantially colder temperature at $\beta = 100$. With no more samples than were used in Figures~\ref{fig:spec} and \ref{fig:rt}, the Euclidean correlator is determined to far higher relative precision. The real-time correlator is well constrained for a full period.

\begin{figure*}
	\centering
	\includegraphics[width=0.48\linewidth]{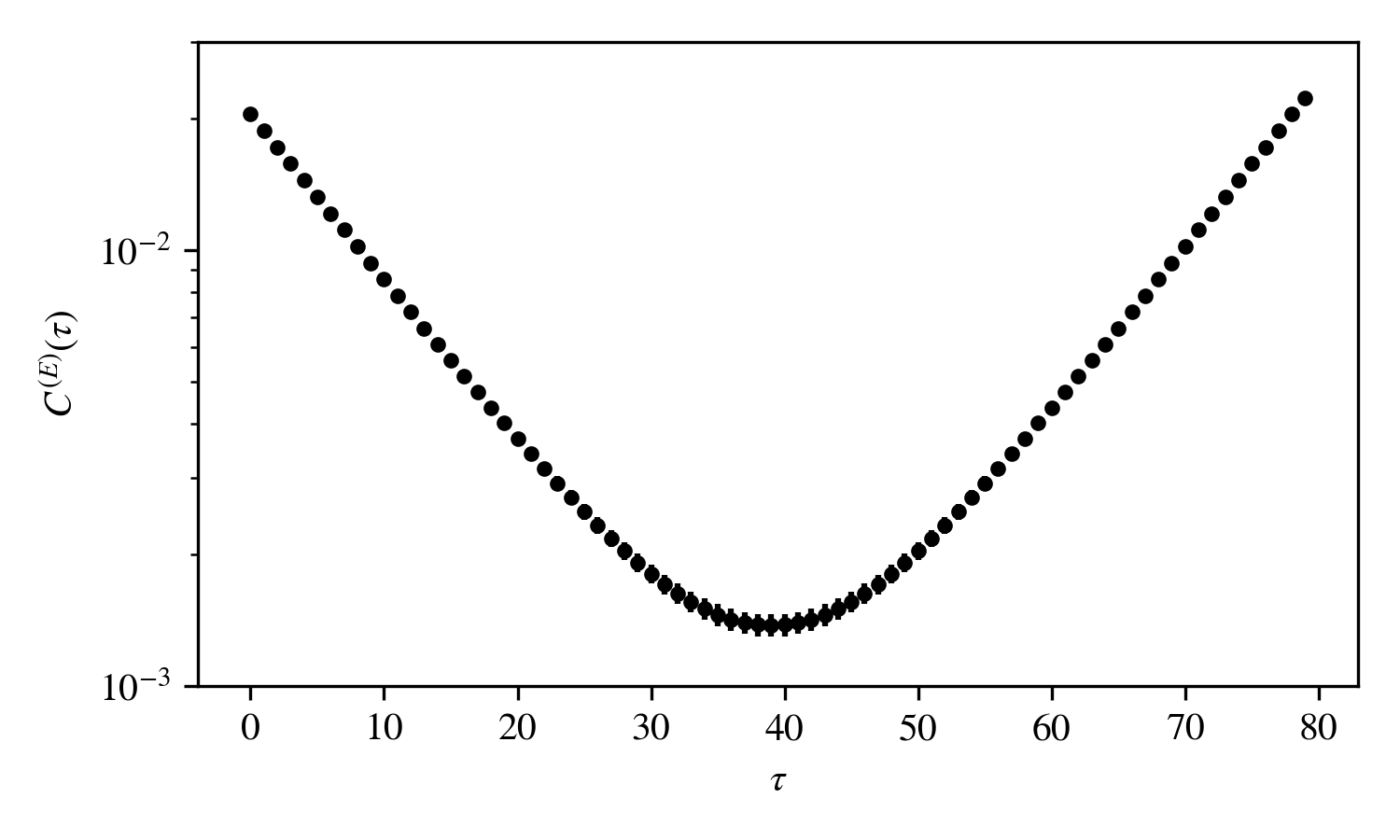}
	\hfill
	\includegraphics[width=0.48\linewidth]{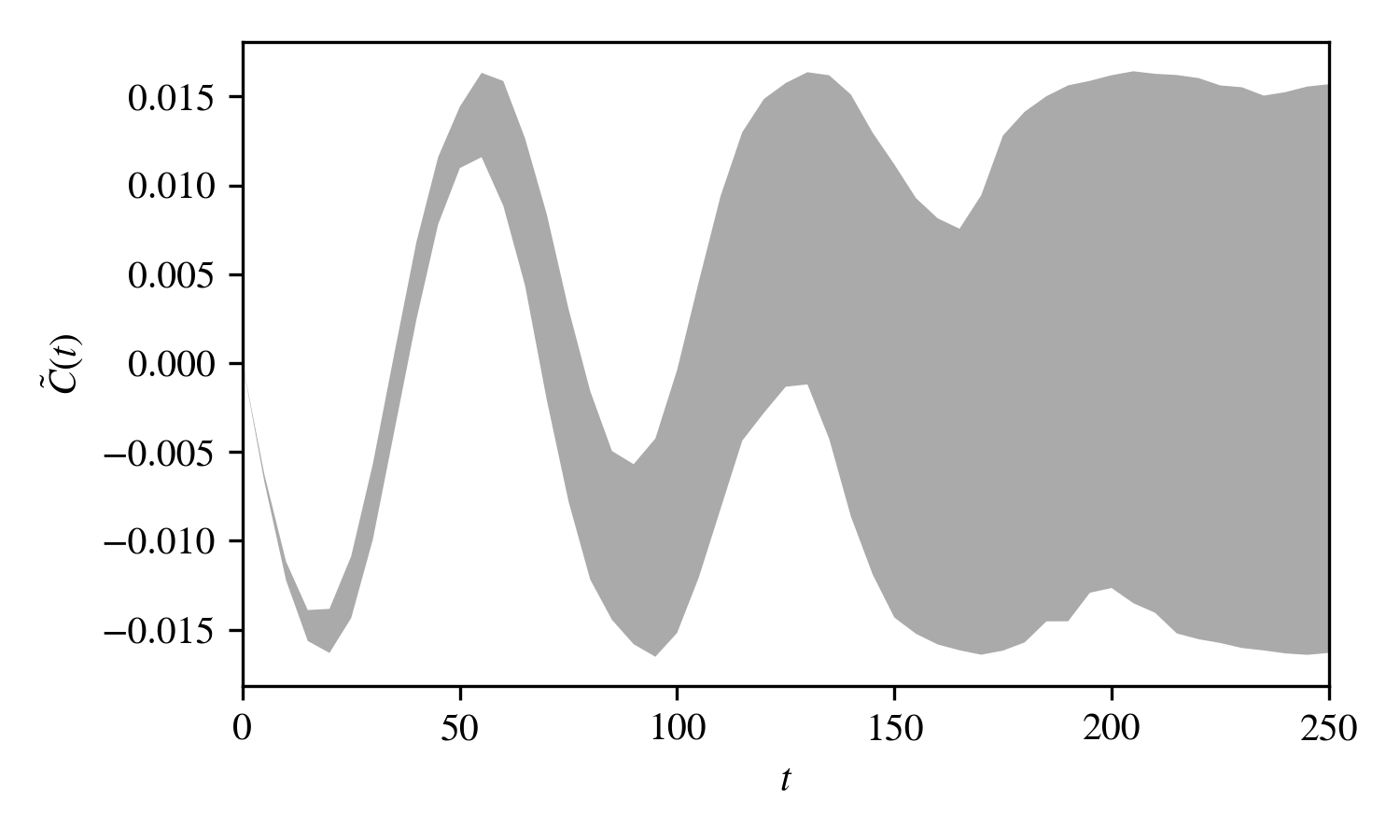}
	\caption{The real-time correlator (right panel) of a scalar field in $2+1$ dimensions, on a $16^2$ spatial lattice at $\beta = 80$. A total of $\sim 1.9 \times 10^5$ samples were used to determine the Euclidean correlator (left panel).\label{fig:scalar-cold}}
\end{figure*}

None of this is special to quantum mechanics, as the only input to the inversion routine is the Euclidean correlator, which may be determined quite precisely on large lattices in high dimension. Figure~\ref{fig:scalar-cold} shows the bounds that may be obtained on the smeared equivalent of $\langle \bar \phi(t) \bar \phi\rangle$, in scalar field theory in $2+1$ dimensions. The Euclidean lattice geometry is $16^2 \times 80$, and $\sim 1.9 \times 10^5$ configurations (albeit not carefully decorrelated) are used to resolve the time-dependence into the second period. The lattice parameters of $m^2 = 0$ and $\lambda = 10^{-2}$ produce an interacting scalar field with a mass of $M \approx 10^{-1}$.

The same real-time correlator, albeit without model-free error estimates, may be obtained much more straightforwardly by fitting the Euclidean correlator (shown in the left panel of Figure~\ref{fig:scalar-cold}) to a single hyperbolic cosine. This fit simultaneously reveals the mass, corresponding to the frequency of the primary sine wive in the real-time correlator, and the amplitude of the same sine wave. That real-time correlators may be determined in such restricted situations is not a novel observation: it is only at high temperatures (or possibly in gapless theories) that the Minkowski two-point function becomes of interest. Indeed, past work on alleviating the real-time sign problem in field theories has focused on high temperatures relative to the time extent~\cite{Alexandru:2016gsd,Alexandru:2017lqr,Lampl:2023xpb}.

\section{Discussion}\label{sec:discussion}
We have described in Sections~\ref{sec:spectral} and \ref{sec:convex} the structure of the space of spectral density functions which are simultaneously non-negative, and consistent with given Euclidean data. This space is convex, enabling standard (and efficient) methods of convex optimization to be used to explore the space. In particular, by passing to the Lagrange dual problem (described in Section~\ref{sec:convex} and derived in Appendix~\ref{app:dual}), we obtain a convex optimization problem whose solution gives a lower (or upper) bound for a desired integral of the spectral density function. As a result, we may efficiently and rigorously estimate smeared versions of the spectral density function or of the real-time correlator; these are demonstrated in Sections~\ref{sec:density} and \ref{sec:realtime} respectively. The real-time correlator is determined with higher precision at lower temperatures, and in Section~\ref{sec:cold} bounds on a real-time correlator in $2+1$ scalar field theory are obtained.

Crucially, the bounds obtained in this manner are information-theoretically tight. To be precise, suppose we are estimating some integral $I = \int d\omega\, \rho(\omega) \mathcal K(\omega)$. The procedure described in this work gives upper and lower bounds, which we denote $I_<$ and $I_>$. Then for any real number $I'$ in the interval $[I_<, I_>]$, there exists a spectral density function $\rho'(\omega)$ simultaneously satisfying:
\begin{enumerate}
	\item The spectral density is positive: $\rho'(\omega) \ge 0$
	\item The spectral density reproduces the integral $I'$: $\int d\omega\, \rho'(\omega) \mathcal K(\omega) = I'$, and
	\item The spectral density function reproduces the Euclidean data, in the sense that the statistic $F$ defined by Eq.~(\ref{eq:statistic}) lies within the $99\%$ confidence interval.
\end{enumerate}
Therefore any further constraints on the value of $I$ must come from additional information, beyond what is available in the given Euclidean data and assumption of positivity.

The above ``no-go theorem'' has one particularly noticeable loophole\footnote{Loopholes are a desirable property in a no-go theorem.}: there is no guarantee that the chosen statistic $F$ yields the tightest possible bounds on any given integral. Any similarly constructed statistic will still yield a convex optimization problem leading to valid bounds, and different statistics may be used for different choices of $\mathcal K$. A somewhat natural alternative is to set the matrix $M$ to be proportional to the identity, thus disregarding the covariance matrix---empirically this resulted in far looser bounds on real-time evolution. The task of searching for superior statistics is left for future work.

All results in this paper have been limited to spectral density functions related to correlators of the form $\langle \mathcal O \mathcal O\rangle$. The general two-point function may involve two \emph{different} operators, in which case the corresponding spectral density function is no longer guaranteed to be non-negative. However, positivity bounds are still available. Selecting a basis of operators $\{\mathcal O_i\}$, the general two-point function may be thought of as a single matrix-valued object $G_{ij}(t) = \langle \mathcal O_i(t) \mathcal O_j(0)\rangle$, with a corresponding matrix-valued spectral density now guaranteed to be positive semi-definite at any frequency $\omega$.

Recent work on improving the signal-to-noise ratio of lattice Monte Carlo calculations has made use of the existence of a set of exact linear relations between different expectation values, to construct control variates~\cite{Bhattacharya:2023pxx,Bedaque:2023ovz,Lawrence:2024xsi}. These same relations can be re-written as linear constraints on the (matrix-valued) spectral density. Nontrivial Schwinger-Dyson relations are not available in scalar field theory when only one operator is considered, but future work analyzing a multi-operator spectral density may be able to make use of these.

A major motivation for the study of real-time dynamics is the goal of determining transport coefficients of strongly interacting field theories from lattice calculations. Those transport coefficients that are related to dissipative processes are typically defined in the long-time limit. It is apparent from the results in the preceding sections that obtaining reliable long-time extrapolations from noisy, discrete lattice data will be difficult, and perhaps not possible in practice, without the imposition of further constraints on the spectral density.

One form of modelling of the spectral density is commonplace in lattice field theory: determining masses in the low-lying spectral excited by some operator or basis of operators. The standard methods for accomplishing this perform, in one way or another, a fit of multiple exponential decays to the Euclidean correlator~\cite{Michael:1982gb,Blossier:2009kd}. This method struggles when it is difficult to model or constrain the impact of states not included in the fit. It may be that including the requirement $\rho(\omega) \ge 0$---which is explicitly imposed on the modelled exponential decays, but not on the unmodelled states---allows the contribution of the nuisance states to be rigorously bounded. A hint in this direction is provided by recent work inspired by the Lanczos method on evading the signal-to-noise problem~\cite{Wagman:2024rid,Hackett:2024xnx}, where the best-fit spectral density is found to violate positivity (indicating that imposing positivity would yield stronger constraints). 
Nevertheless it is unclear how the methods of this paper might best be brought to bear on the problem of determining masses: absent additional assumptions, the mass of a particle is not a convex function of the spectral density.

Finally, note that the algorithm to solve the dual optimization problem may be considered as an automated theorem prover. Given a point in the space of Lagrange multipliers, we may write down a proof of some bound on the integral $\int \mathcal K \rho$ being studied. Such a proof is not guaranteed to be valid---there may be incorrect steps. The proof will be valid precisely when the chosen values of the Lagrange multipliers correspond to a (dual-)feasible point. Therefore, the space of proofs, viewed as a vector space isomorphic to the vector space of Lagrange multipliers, has a convex subset corresponding to the space of valid proofs. The strength of the bound is a convex function on this space, and solving the dual problem is equivalent to searching for the strongest theorem which can be established with a proof of this form.

\begin{acknowledgments}
	This work was preceded and informed by many insightful conversations with Tanmoy Bhattacharya, Tom Cohen, Tom DeGrand, Brian McPeak, Duff Neill, and Paul Romatschke. Evan Berkowitz provided invaluable comments on an earlier draft.

	This work was supported by a Richard P.~Feynman fellowship from the LANL LDRD program. Los Alamos National Laboratory is operated by Triad National Security, LLC, for the National Nuclear Security Administration of U.S. Department of Energy (Contract Nr. 89233218CNA000001).
\end{acknowledgments}

\appendix
\section{Ill-posedness}\label{app:ill}

It is often said that the primary obstacle to obtaining real-time information from an imaginary-time simulation is that the inversion is ``ill-posed''. The same obstacle appears to obstruct the determination of many other integrals of the spectral density. The mathematical literature provides a rigorous definition of the notions of well- and ill-posed problems, but it turns out that these do not map well onto the intuition that guides statements about the difficulty of spectral reconstruction. The discussion below is intended to help clarify the situation.

To keep the discussion from getting too abstract, here are five computations one might wish to perform on a Euclidean correlator $G^{(E)}(\tau)$, given data at discrete $\tau_i$ equipped with statistical errors:
\begin{enumerate}[(i)]
	\item Compute a sum of the correlator over time separations: $\Sigma = \sum_i G(\tau_i)$
	\item Compute an integral over all times, yielding a susceptibility: $\chi = \int_0^\beta d\tau\,G^{(E)}(\tau)$
	\item Compute the smeared spectral density function; for concreteness let us consider the particular integral $R = \int_0^\infty \rho(\omega) e^{-(\omega-1)^2}\,d\omega$
	\item Compute the real-time correlator at some finite time separation $t=1$, without smearing: $G(1) = \int_0^\infty \rho(\omega) e^{-i \omega}$
	\item Compute the spectral density at a fixed energy $\omega=1$, without smearing: $\rho(1)$.
\end{enumerate}
Intuitively, the first two tasks are easy (certainly they are done on a regular basis in lattice field theory), and the last two are not. The middle task, along with its variations, has been the subject of this work.

 The notion of an ill-posed problem appears most often in the study of PDEs, and in that context a standard definition is provided (quoting~\cite{pcm:pde}):
\setcounter{definition}{-1}
\begin{definition}[Ill-posed problem for a PDE]
	A given problem for a PDE is said to be well-posed if both existence and uniqueness of solutions can be established for arbitrary data that belongs to a specified large space of functions, which includes the class of smooth functions. Moreover, the solutions must depend continuously on the data. A problem that is not well-posed is called ill-posed.
\end{definition}
This definition provides three ways in which a problem might be ill-posed: a solution may sometimes not exist, the solution may not be uniquely determined, or the solution might not depend continuously on the data. For the spectral reconstruction problem it makes sense to assert that only functions $G^{(E)}(\tau)$ originating from some non-negative spectral density will ever be considered (obviating the question of existence). Indeed, without the restriction to non-negative spectral densities, all tasks except the first are ill-posed, in the sense that a spectral density function can be found assigning any real numbers to $\chi$, $R$, $G(1)$, and $\rho(1)$. Moreover no meaningful bounds can be obtained for any of these tasks.

Dropping the question of continuity, which is plainly secondary to uniqueness\footnote{Although one might reasonably ask about the continuity of the upper and lower bounds, as functions of the provided data.}, we might define a ill-posed reconstruction problem as follows.
\begin{definition}[Ill-posed spectral reconstruction]
	A spectral reconstruction problem---a request for $I \equiv \int \mathcal K \rho$ given $C_i = \int K_i \rho$---is ill-posed if $I$ cannot be uniquely determined from $C_i$. That is, it is ill-posed if there are two spectral densities $\rho,\rho' \ge 0$ such that $\int K_i \rho = \int K_i \rho'$ but $\int \mathcal K \rho \ne \int \mathcal K \rho'$.
\end{definition}
Applying this definition we find that once again the spectral reconstructions problems (ii-v) are all ill-posed. The case of task (ii) makes it plain that the fact that a spectral reconstruction problem is ill-posed does not imply that no useful information can be gained about the desired quantity in practice, or even that determining that information must be difficult.

We can subdivide the set of ill-posed problems further by considering various stronger statements. For example, in the case of problem (v), the correlators $C_i$ provide no information at all about the spectral density $\rho(1)$, in the sense that no finite bounds on $\rho(1)$ can be proven. This is not true of any of the other four problems. However, while the correlator $G(t)$ has finite bounds, these come only from the Euclidean correlator at zero time separation:
\begin{equation}
	|G(t)| \le G^{(E)}(\tau = 0)
\end{equation}
No finite amount of information regarding the Euclidean correlator enables one to tighten pointwise bounds on the real-time correlator further. This separates problems (i-iii), for which arbitrarily tight bounds can be achieved given sufficient information, from problems (iv) and (v).

Note that the difficulties in performing various reconstruction tasks are not due to computational complexity, but a lack of information. Since bounds are obtained by convex optimization, optimal bounds can be computed efficiently.

 We conclude by noting the following empirical statement about the difficulty of determining the smeared real-time correlator, which has little in common with the question of whether a problem is well-posed:
\begin{conjecture*}
	For any $\sigma > 0$, there exists some $A,t_0,\epsilon > 0$ such that, in order for the bound on $\tilde G_\sigma(t)$ to have size less than $\epsilon$, at least $A e^{t/t_0}$ samples are needed (regardless of the chosen lattice spacing).
\end{conjecture*}

\section{Interior-point method}\label{app:ipm}
\begin{figure}[b]
\begin{algorithm}[H]
	\begin{algorithmic}[1]
		\Require $x$ strictly feasible, $\mu > 1$, $t_0 > 0$, $\epsilon > 0$
		\State $t \gets t_0$
		\While{$m/t < \epsilon$}
		\State $x^*(t) \equiv \arg\min_{x'} f(x') + \phi(x')/t$
		\Statex (Minimize by gradient descent, starting at $x$)
		\State $x \gets x^*(t)$
		\State $t \gets \mu t$
		\EndWhile
		\State\Return $x$
	\end{algorithmic}
	\caption{Interior-point convex optimization\label{alg:ipm}}
\end{algorithm}
\end{figure}

The algorithm used to solve the convex program (\ref{eq:spectral-dual}) is an interior-point method; this is a standard algorithm and widely implemented in a variety of commercial and open-source optimization packages. However, as noted in the text, there are formally an infinite number of constraints, making it difficult to encode the problem in a form accepted by such software. The purpose of this appendix is to detail the interior-point method used; the associated code has been made available at~\cite{scamp}.

The interior-point method is based around a barrier function: a convex function of the search space which is smooth on the set of feasible points, but diverges to $+\infty$ as the boundary of feasibility is approached. Denoting such a function $\phi(x)$, and the objective function $f(x)$, we may define for any real $t > 0$ a modified objective function:
\begin{equation}
	f_t(x) = f(x) + t^{-1} \phi(x)
	\text.
\end{equation}
This modified objective function also diverges to $+\infty$ near the boundary of the feasible region, for any non-negative $t$. However, for sufficiently large $t$, it is a good approximation to $f$ throughout the interior of the feasible region.

The interior-point method, detailed in Algorithm~\ref{alg:ipm} proceeds by minimizing $f_t(x)$ at successively larger values of $t$. The gradient descent step is implemented by BFGS~\cite{fletcher2000practical}. The performance of the algorithm is in practice not very sensitive to the choice of hyperparameters $\mu$, $t_0$, and $\epsilon$; for all results in this work we have used $\mu = 1.5$, $\epsilon=10^{-10}$, and $t_0 = 10^{-3}$. Note also that the implementation of this algorithm used here uses only $64$-bit floating point numbers, and no evidence of a need for higher precision was found.

This interior-point method assumes that we begin with a known feasible point. Typically this is not the case, and a separate ``phase one'' algorithm must run in order to find such a feasible point. One way to find such a feasible point, if it exists, is to solve a convex program of the form
\begin{equation}
	\begin{split}
		\text{maximize }&s\\
		\text{subject to }&h_i(x) \ge s
		\text.
	\end{split}
\end{equation}
Here the functions $h_i$ define all constraints on the original convex program. Finding a feasible point of this convex program is trivial (one simply selects $x$ at random and then computes a sufficient value of $s$). The optimization of this program proceeds by the interior-point method described above and may terminate as soon as $s \ge 0$, since any feasible point will do. In practice this happens quite early in the optimization, and so the vast majority of time spent by the solver is on the second phase.

It remains to define the barrier function $\phi(\mu,\ell)$ used to implement the constraints of (\ref{eq:spectral-dual}). It is no difficult matter to write down a function with the desired properties:
\begin{equation}
	\phi(\mu,\ell) = -\log \mu - \int_0^\infty \log \lambda(\omega;\ell) \,d\omega
	\text.
\end{equation}
As the second term is an integral of a smooth function of one variable, it may be efficiently and reliably evaluated without difficulty. In principle this can be done adaptively, using the value and derivative at $\omega_k$ to determine the next energy, $\omega_{k+1}$, at which the function must be probed. It is sufficient in practice to simply pick a dense set of points on which to approximate the integral via a Riemann sum or trapezoid rule. Note that the descent algorithm scales linearly with the number of points at which $\lambda(\omega)$ is evaluated. In contrast, when solving the primal problem, the scaling is roughly cubic with the number of points at which $\rho(\omega)$ is to be evaluated.

\section{Dualization}\label{app:dual}
In this appendix we derive an appropriate Lagrange dual of the (infinite-dimensional) primal problem:
\begin{equation}
	\begin{split}
		\text{minimize }&\int_0^\infty \mathcal K(\omega) \rho(\omega) d\omega\\
		\text{subject to }&\rho(\omega) \ge 0\\
		\text{and }&v(\rho)^T M v(\rho) \le 1\\
		\text{where }&v_i(\rho) \equiv C_i - \int_0^\infty K(\tau_i,\omega) \rho(\omega) d\omega\text.
	\end{split}
\end{equation}
Above, the functions $\mathcal K(\omega)$ and $K(\tau,\omega)$, together with the matrix $M$ and the measured correlators $C$, define the optimization program. The optimization is performed over the space of spectral density functions $\rho(\omega)$.

The dualization begins by defining a Lagrange functional. With two inequality constraints, we have two Lagrange multipliers: a function $\lambda(\omega)$ to enforce the positivity of $\rho$, and a real number $\mu$ to confine us to the confidence region. The Lagrange functional reads
\begin{equation}
	\begin{split}
	L(\rho;\lambda,\mu)
	=
		&\int \mathcal K(\omega) \rho(\omega) d\omega\\
		&- \int \lambda(\omega) \rho(\omega) d\omega - \mu (1 - v^T M v)
	\text.
	\end{split}
\end{equation}
Here, as in the body of the paper, the vector $v$ is a function of $\rho$, defined by
\begin{equation}
	v_i(\rho) \equiv C_i - \int K(\tau_i,\omega) \rho(\omega) d\omega
	\text.
\end{equation}
It is straightforward to confirm that the primal optimum is equal to
\begin{equation}
	p^* = \min_{\rho} \max_{\lambda,\mu \ge 0} L(\rho;\lambda,\mu)
	\text.
\end{equation}
Moreover, evaluating the inner maximization yields precisely the primal optimization problem written above.

The dual problem is defined by swapping the order of the optimizations, so that the dual optimum is equal to
\begin{equation}\label{eq:dual-optimum}
	d^* = \max_{\lambda,\mu \ge 0} \min_\rho L(\rho;\lambda,\mu)
	\text.
\end{equation}
We must now evaluate the inner minimization, to obtain an optimization problem over the space of $(\lambda,\mu)$. For both brevity and clarity, we may treat functions of $\omega$ as vectors. Expanding the definition of $v$, the Lagrange function now reads
\begin{equation}
	\begin{split}
		L(\rho;\lambda,\mu) ={}& (\mathcal K - \lambda - 2 \mu K^T M^T C )^T \rho\\
		&+ \mu \rho^T K^T M K \rho
 + \mu (C^T M C - 1)	\text.
	\end{split}
\end{equation}
The Lagrange dual function $g(\lambda,\mu)$ is defined by performing the inner minimization in Eq.~(\ref{eq:dual-optimum}). The matrix $M$ is positive definite, but $K^T M K$ is not full-rank. When the vector $(\mathcal K - \lambda - 2\mu K^T M^T C)$ is orthogonal to the null space of $K^T M K$, the minimum is finite; otherwise the Lagrange function is unbounded below. The outer maximization therefore imposes a restriction to the space orthogonal to the null space. In this space we may parameterize $\lambda$ by a vector $\ell$:
\begin{equation}
	\lambda = \mathcal K - K^T \ell\text.
\end{equation}
Persisting with this projection onto the row space of $K$, the Lagrange function is
\begin{equation}
	L(\rho;\ell,\mu) =\ell^T K \rho + \mu (K \rho - C)^T M (K \rho - C) - \mu \text.
\end{equation}
We may not assume that $\rho$ itself lies in the row space of $K$, but decomposing it into $\rho = \rho_\parallel + \rho_\perp$, with $\rho_\perp$ in the null space and $\rho_\parallel$ in the row space of $K$, we see that the Lagrange function is in fact independent of $\rho_\perp$.
Analogous to what we did for $\lambda$, we next decompose $\rho_\parallel = K^T r$, finally writing the Lagrange function as
\begin{equation}
	L(r;\ell,\mu) = \ell^T \mathfrak K r + \mu (\mathfrak K r - C)^T M (\mathfrak K r - C) - \mu\text.
\end{equation}
Above we have defined the symmetric matrix $\mathfrak K = K K^T$.
For fixed $(\ell,\mu)$, the Lagrange function is minimized at $r=r^*$, where
\begin{equation}
	\mathfrak K r = -\frac{1}{2\mu} M^{-1} \ell + C
	\text.
\end{equation}
Therefore, performing the minimization over $r$, we obtain the Lagrange dual function (restricted to the space on which it is finite):
\begin{equation}
	g(\ell,\mu) =
	-\mu
	- \frac{1}{4\mu} \ell^T M^{-1} \ell 
	+ \ell^T C
	\text.
\end{equation}
We conclude that the dual optimization problem may be written
\begin{equation}
	\begin{split}
		\text{maximize }&\ell^T C - \frac{1}{4\mu} \ell^T M^{-1} \ell - \mu\\
		\text{subject to }& \mathcal K - K^T \ell \ge 0\\
		\text{and }&\mu \ge 0
	\text.
	\end{split}
\end{equation}
Note that the objective function is convex.

\bibliography{refs}
\end{document}